\documentclass{aa}
\usepackage{txfonts}
\usepackage{amssymb}
\usepackage{natbib}
\usepackage{graphicx}
\begin{document}
\title{\textbf{\object{XRF~050406} late-time flattening: an inverse Compton component?}}
\author{A. Corsi\inst{1,2,3} \and L. Piro \inst{1}}
\institute{IASF-Roma/INAF, Via Fosso del Cavaliere 100, 00133 Roma, Italy.
\and Universit\`a degli studi di Roma ``La Sapienza'', Piazzale Aldo Moro 5, 00185 Roma, Italy.
\and INFN - Sezione di Roma c/o Dip. di Fisica - Universit\`a degli studi di Roma ``La Sapienza'', Piazzale Aldo Moro 5, 00185 Roma, Italy.}
\offprints{A. Corsi -- Alessandra.Corsi@iasf-roma.inaf.it}
\date{}
\abstract{}{We investigate for possible evidence of inverse Compton (IC) emission in the X-ray afterglow of \object{XRF~050406}.}{In the framework of the standard fireball model, we show how the late-time flattening observed in the X-ray light curve between $\sim 10^{4}$~s and $\sim 10^{6}$~s can be explained in a synchrotron-plus-IC scenario when the IC peak frequency crosses the X-ray band.} {We thus conclude that the appearance of an IC component above the synchrotron one at late times successfully accounts for these X-ray observations.}{}
\keywords{Gamma rays: bursts -- X-rays: Individuals (\object{XRF~050406}) -- X-rays: bursts -- radiation mechanisms: non-thermal}
\authorrunning{}
\titlerunning{\object{XRF~050406}: an IC component?}
\maketitle
\section{Introduction}
The ``Burst Alert Telescope'' \citep[BAT;][]{Barthelmy2005} on board Swift \citep{Gehrels2004} was triggered by \object{GRB~050406} on April 6, 2005, at 15:58:48.40~UT \citep{Parsons2005}. The BAT located the burst at RA=$02^{\rm{h}}17^{\rm{m}}53^{\rm{s}}$ and Dec=$-50^{\circ}10'52''$ (J2000), with an uncertainty of $3$~arcmin \citep{Krimm2005}. The photon index of the $15-350$~keV time-averaged spectrum was $\Gamma=2.38\pm0.34$ \citep{Krimm2005}. Because of the spectral softness, this burst was classified as an X-ray flash \citep[XRF;][]{Heise2001}. The $15-350$~keV fluence was $(1.0^{+1.13}_{-0.36})\times10^{-7}$~ergs~cm$^{-2}$ \citep{Romano2005}. Assuming a redshift of $z=2.44$ \citep{Shady2006}, the isotropic energy release was $\sim1.4\times10^{51}$~ergs.  

The ``X-Ray Telescope'' \citep[XRT;][]{Burrows2005a} imaged the BAT field starting from $84$~s after the trigger, and the X-ray counterpart of \object{XRF~050406} was found during the on-ground analysis \citep{Cusumano2005,Capalbi2005}. The ``Ultra-Violet/Optical Telescope'' \citep[UVOT; ][]{Roming2005} also started imaging about $88$~s after the trigger. The optical afterglow was not detected on-board \citep{Landsman2005}, but subsequent on-ground analysis revealed a source within the XRT error circle \citep{Rol2005}. Late-time observations ($\sim 7.8$~hr after the burst) performed by the Magellan/Clay Telescope revealed a single faint source ($R=22.0\pm0.09$) located at RA=$02^{\rm{h}}17^{\rm{m}}52^{\rm{s}}.3$ and Dec=$-50^{\circ}11'15''$ \citep[J2000;][]{Berger2005b,Berger2005a}.

The X-ray light curve of \object{XRF~050406} shows some very interesting features. In addition to an X-ray flare \citep{Burrows2005b}, it is characterized by a flattening at late times ($t~\gtrsim4200$~s from the trigger). In this work we test whether the flattening can be related to the appearance of an IC component \citep[e.g., ][]{Wei1998,Panaitescu2000,Wei2000,SariEsin2001}, in the context of the standard fireball model \citep[e.g. ][]{Sari1998}. In Sect. \ref{sec1} we summarize the observed properties of the broad-band afterglow. We refer to the $0.2-10$~keV light curve and spectral analysis presented by \citet{Romano2005} and to Table 1 in \citet{Shady2006} for the optical data. In Sect. \ref{sec2} we constrain the parameters of the fireball in a synchrotron-plus-IC scenario and in Sect. \ref{sec4} we present our conclusions.

Hereafter, $F(\nu,t)\propto t^{-\alpha} \nu^{-\beta}$ is the flux density at the observer's time $t$ and observed frequency $\nu$; $\alpha$ and $\beta$ are the temporal and spectral indices respectively; $E_{52}=E/(10^{52}$~ergs) is the isotropic equivalent energy of the shock in units of $10^{52}$~ergs; $n$ is the particle number density of the ambient medium in units of $1$~particle/cm$^{3}$; $\epsilon_{B,-2}=\epsilon_{B}/10^{-2}$ is the fraction of the shock energy, which goes in magnetic energy density behind the shock, in units of $10^{-2}$; $\epsilon_{e,0.5}=\epsilon_{e}/0.5$ is the fraction of shock energy that goes into accelerating the electrons, in units of $0.5$; $p$ is the power-law index of the electron energy distribution \citep{Sari1998}. We adopt $z=2.44$ for the redshift of the source \citep{Shady2006}.
\begin{table*}
\begin{minipage}[t]{\textwidth}
\begin{center}
\caption{Closure relationships between the X-ray afterglow spectral and temporal indices ($\beta=1.1\pm0.3$, $\alpha_{1}=1.58\pm0.17$) of \object{XRF~050406} in the standard synchrotron fireball model for a wind or ISM environment. In parenthesis: relationships in an ISM modified for the effect of IC cooling. The (u) marks those relationships that are not affected by IC emission.} \label{tab:relazioni_chiusura} 
\renewcommand{\footnoterule}{}
\begin{tabular}{c c c c c c}
\hline\hline
&&\multicolumn{2}{c}{ISM environment}&\multicolumn{2}{c}{Wind environment}\\
\hline
&& Expected relation& Observed value&Expected relation& Observed value\\
\hline
a)&$\nu_{c}<\nu_{X}<\nu_{m}$ & $2\alpha-\beta=0$ (u)& $2.1\pm0.4$& $2\alpha+\beta-1=0$& $3.3\pm0.5$\\
\hline
b)&$\nu_{c}<\nu_{m}<\nu_{X}$ & $2\alpha-3\beta+1=0$ (u)& $0.86\pm0.96$ & $2\alpha-3\beta+1=0$& $0.86\pm0.96$\\
\hline
c)&$\nu_{m}<\nu_{X}<\nu_{c}$ & $2\alpha-3\beta=0$ (u)& $-0.14\pm0.96$ & $2\alpha-3\beta-1=0$& $-1.1\pm1.0$\\
\hline
d)&$\nu_{m}<\nu_{c}<\nu_{X}$ & $2\alpha-3\beta+1=0$~$(\alpha+\frac{3\beta^{2}-6\beta+1}{2(2-\beta)}=0)$ & $0.86\pm0.96$~$(0.48\pm0.31)$ & $2\alpha-3\beta+1=0$& $0.86\pm0.96$\\
\hline
\end{tabular}
\end{center}
\end{minipage}
\end{table*}

\section{The optical-to-X-ray afterglow}
\label{sec1} The $0.2-10$~keV light curve of \object{XRF~050406} shows complex behavior \citep{Burrows2005b}, with a power-law decay underlying a flare peaking at about $210$~s after the trigger and a flattening between $\sim 10^{4}$~s and $\sim 10^{6}$~s. Excluding the flare, a broken power-law model with $\alpha_{1}=1.58^{+0.18}_{-0.16}$, $\alpha_{2}=0.50\pm0.14$, and $t_{\rm{break}}\sim4200$~s yields a good fit to the data \citep{Romano2005}.

 The mean X-ray energy index was $\beta=1.1\pm0.3$ during the first $600$~s after the trigger and $\beta=1.06\pm0.24$ between $\sim 600$~s and $\sim 2\times10^{4}$~s \citep{Romano2005}. These results were obtained by fitting the data with an absorbed power-law model with hydrogen column density ($N_{\rm{H}}$) fixed to the Galactic value of $2.8\times10^{20}$~cm$^{-2}$ \citep{Dickey1990}. The $3\sigma$ upper-limit for the total (Galactic plus intrinsic) $N_{\rm{H}}$ along the line of sight is $9 \times 10^{20}$~cm$^{-2}$
\citep{Romano2005}.

The first UVOT observations were performed in the $V$-band, between $113$~s and $173$~s. The highest flux was measured in a $50$~s long exposure, starting $113$~s after the trigger. During this observation, the measured magnitude was $18.92\pm0.31$~mag \citep{Shady2006}; after correction for Galactic extinction \citep[$E(B-V)=0.022$ mag and $A_{V}/E(B-V)=3.1$, ][]{Schlegel}, this corresponds to a $V$-band flux of $0.11\pm0.04$~mJy. Since at about $100$~s the $0.2$~keV flux was $\sim 0.05$~mJy \citep[using a conversion factor of $6.5\times10^{-11}$~erg~cm$^{-2}$~count$^{-1}$ and a spectral index of $\beta\sim 1.1$, as found by ][]{Romano2005}, the observed optical-to-X-ray spectral index at $100$~s, $\beta_{opt-X}\sim 0.15$, turned out to be extremely flat. Such a flat value was not found in the analysis performed by \citet{Shady2006}, where the mean optical and X-ray flux between $220$~s and $950$~s were considered.

\section{Synchrotron plus IC model}
\label{sec2} 
As seen in the previous section, the X-ray light curve of XRF~050406 is characterized by a steep decay followed by a flattening after $t\sim 4200$~s (see e.g. Fig. \ref{syn}). Recently, \citet{Chincarini2005} and \citet{Nousek2005} found that the X-ray light curves of Swift GRBs usually present an initial steep decay ($t \lesssim 500$~s), followed by a shallower one and, finally, by
further steepening. In the case of \object{XRF~050406}, the temporal decay observed before the break ($\alpha_{1} \sim 1.58$) is not as steep as usual \citep[$3\leq\alpha\leq5$, ][]{Tagliaferri2005} and the curvature relation $\alpha=\beta+2$ is not satisfied. These arguments suggest that the emission preceding the late-time flattening could be part of the afterglow \citep{Romano2005} rather than the tail of the prompt emission. In this scenario, the superimposed flare could be interpreted in the framework of the late internal-shock model \citep{Burrows2005b,Fan2005,Wu2005,Romano2005}. On the other hand, if the flare marks the onset of the afterglow \citep{Piro2005,Galli2005}, the analysis presented here is valid for $t > 210$~s.

Several authors have considered the effect of IC emission in GRB X-ray light curves at late times \citep[e.g. ][]{Wei2000,SariEsin2001}. In this section we test a model in which the steep part of the X-ray light curve is produced by synchrotron emission while the late-time flattening is due to the appearance of an IC component at that point in time. 
\subsection{Synchrotron component}
\label{sec3.1} 
In Table \ref{tab:relazioni_chiusura} we summarize the closure relationships between the spectral index $\beta$ and the early temporal decay index $\alpha_{1}$ for the X-ray afterglow of \object{XRF~050406}. The relation
\begin{equation}
\nu_{m}=2\times10^{13}~\frac{f(p)}{f(2.5)}~(1+z)^{1/2}\epsilon_{B,-2}^{1/2}\epsilon_{e,0.5}^{2}E_{52}^{1/2}t^{-3/2}_{\rm{d}}~{\rm Hz}
\end{equation} is the injection frequency, with $f(p)=\left(\frac{p-2}{p-1}\right)^{2}$, and
\begin{equation}
\nu_{c}=2.7\times10^{15}~(1+z)^{-1/2}\epsilon_{B,-2}^{-3/2}E_{52}^{-1/2}n^{-1}t^{-1/2}_{\rm{d}}(1+x)^{-2}~{\rm Hz}
\end{equation} 
is the cooling frequency \citep{SariEsin2001}; $t_{\rm{d}}$ is the observer's time in units of days; $x$ is the ratio of the IC to synchrotron luminosity, i.e. $x \sim \sqrt{\epsilon_{e}/\epsilon_{B}}$ in the fast-cooling regime and $x \sim \sqrt{\epsilon_{e}/\epsilon_{B}} \times (t/t^{0}_{\rm{IC}})^{-(p-2)/(2(4-p))}$ in the slow-cooling IC-dominated regime \footnote{These expressions for $x$ are valid if $x^{2} >> 1$  \citep{SariEsin2001}.}\citep{SariEsin2001}; $t^{0}_{\rm{IC}}$ is the time at which the transition from slow cooling to fast cooling takes place (when $\nu_{m}$ equals $\nu_{c}$). It is important to note that if IC is an efficient cooling mechanism, the transition from fast cooling to slow cooling is delayed by a factor of $\epsilon_{e}/\epsilon_{B}$ \citep{SariEsin2001}.

In the case of a constant density interstellar medium (ISM), we indicate in parenthesis the closure relationships modified for the inclusion of IC emission. In particular in scenario d), the same relation of case b) holds, when considering only synchrotron emission. However, the addition of IC modifies the expression of $\nu_{c}$ for a factor of $(1+x)^{-2}$. Since during the slow cooling regime, $x$ evolves with time, the temporal decay index at $\nu \ge \nu_{c}$ changes from $3/4(p-1)+1/4$ to $3/4(p-1)+1/4-(p-2)/(8-2p)$ \citep{SariEsin2001,Corsi2005} and the corresponding closure relationship changes as indicated in parenthesis in Table \ref{tab:relazioni_chiusura}. In scenario c), since the flux at $\nu \le \nu_{c}$ does not depend on the expression of $\nu_{c}$, the closure relationship is not modified. Finally, in the fast-cooling regime, $x$ does not depend on time, thus the temporal decay index at frequencies above $\nu_{c}$ does not change (cases a) and b)). 

On the basis of the closure relationships, scenarios b), c), and d) are compatible with the observations in both an ISM and a wind environment \citep{Chevalier1999}. Hereafter, we will limit our discussion to the ISM case.

\begin{figure}
\centering \resizebox{\hsize}{!}{\includegraphics{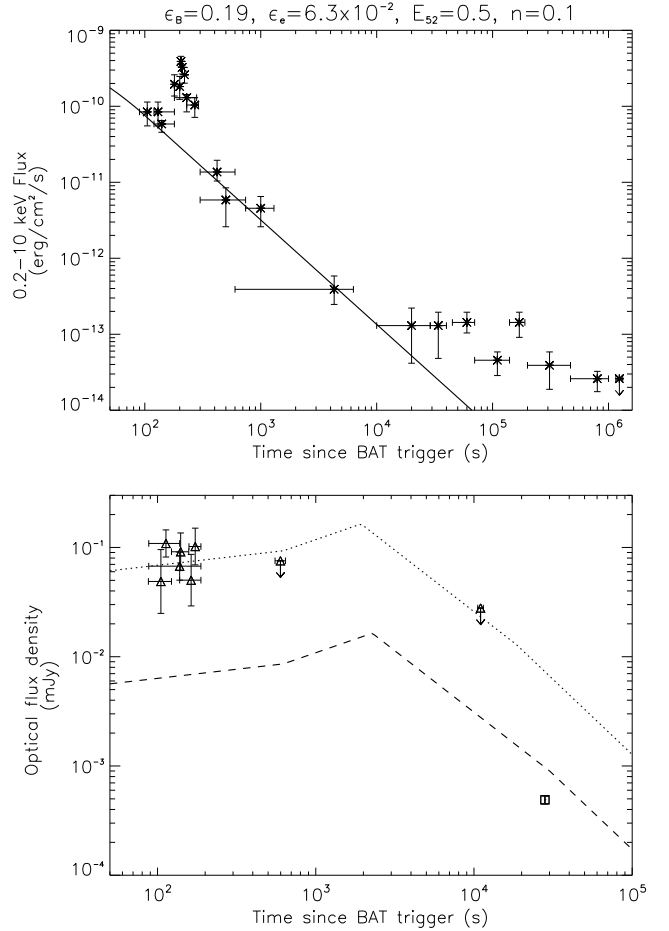}}
\caption{$0.2-10$~keV (upper panel, solid line), $V$-band (lower panel, dotted line), and $R$-band (lower panel, dashed line) light curves in a standard synchrotron fireball model with $\epsilon_{B}=0.19$, $\epsilon_{e}=6.3\times10^{-2}$, $E_{52}=0.5$, $n=0.1$, $p=2.5$, so as to fit the observed optical and X-ray fluxes at $100$ s, i.e. to satisfy the conditions $\nu_{m}$(100~s)~$\sim 0.2$~keV, $\nu_{c}$(100~s)~$\sim 0.03$~keV (see text). The $0.2-10$ keV data points (upper panel, crosses), the $V$-band data points (lower panel, triangles), and the the $R$-band data point (lower panel, box) are taken from \citet{Romano2005}, \citet{Shady2006}, and \citet{Berger2005b}, respectively. The $R$-band light curve and data point have been shifted down by a factor of 10 for clarity.} 
\label{syn}
\end{figure}

We have seen in Sect. \ref{sec1} that the optical-to-X-ray spectral index before the flare is rather flat. In a standard synchrotron scenario, such a flat value cannot be explained unless the synchrotron peak frequency is between the optical and the X-ray band. In Fig. \ref{syn} we show that setting $\nu_{c}$(100~s)~$\sim 0.03$~keV (peak frequency between the optical and the X-ray band) and $\nu_{m}($100~s)~$\sim0.2$~keV (case b) in Table \ref{tab:relazioni_chiusura}), marginal consistency with the data can be obtained. However, under these conditions, $\nu_{m}$ crosses the $V$-band around $2 \times 10^{3}$ s; until that time, the light curve (Fig. \ref{syn}) features a rise, while \object{XFR~050406} was no longer visible at $\sim 600$~s above the background \citep{Shady2006}, suggesting a progressive fading of the optical afterglow \citep[the $V$-band upper-limits adopted here are standard $3\sigma$ upper-limits, thus the corresponding limiting fluxes are $3$ times higher than the $1\sigma$ upper-limits reported by ][]{Shady2006}. Note also that in a synchrotron scenario, $\beta \gtrsim 0.5$ at frequencies above the peak one. Since $\nu_{c}$(100~s)~$\sim 0.03$~keV and $\nu_{m}($100~s)~$\sim0.2$~keV imply $\beta=0.5$ for $\nu_{c} \le \nu \le \nu_{m} = 0.2$~keV, in Fig. \ref{syn} we are minimizing the peak-flux value required to fit the optical and X-ray data at $100$~s, thus minimizing the rise in the optical flux observed until the peak frequency is above the optical band.

To account for a \textit{decreasing} optical emission, the synchrotron peak frequency should be below the optical band. In this case, to fit the early X-ray afterglow while \textit{overestimating as little as possible} the optical flux around $\sim 100$~s, the best choice is to set $\nu_{c}$(100~s)$ \le 10^{14}$~Hz  and $\nu_{m}($100~s)~$\sim 0.2$~keV (case b) in Table 1). In fact, then $\beta_{opt-X}=0.5$, which is the flattest value allowed in a standard synchrotron scenario when the peak frequency is below the optical band (and $p > 2$). Under these assumptions, normalizing to the observed X-ray flux at $100$~s, the predicted $V$-band flux is overestimated for a factor of $\sim~5$. In the $R$-band, the predicted light curve will decrease as $t^{-1/4}$ until the time at which $\nu_{m}$ crosses the band ($\sim 2300$~s) and as $t^{-\frac{3}{4}(p-1)-\frac{1}{4}}$ up to the time of the Magellan/Clay observation ($t\sim 3\times10^{4}$~s). Since in this case the measured X-ray spectral and temporal indices imply $\left\langle p\right\rangle=2.5\pm0.3$, then $t^{-\frac{3}{4}(p-1)-\frac{1}{4}}=t^{-1.375}$, and we expect the predicted $R$-band flux to be around a factor of $\sim 2$ above the Magellan/Clay data point. Requiring additional extinction in the GRB host galaxy or a contribution to the early X-ray flux coming from the rising part of the flare (or a combination of these two effects) helps account for the broad-band observations, as we see in the following section.
\subsection{IC component}
\begin{figure*}[!]
\centering
\includegraphics[width=17 cm]{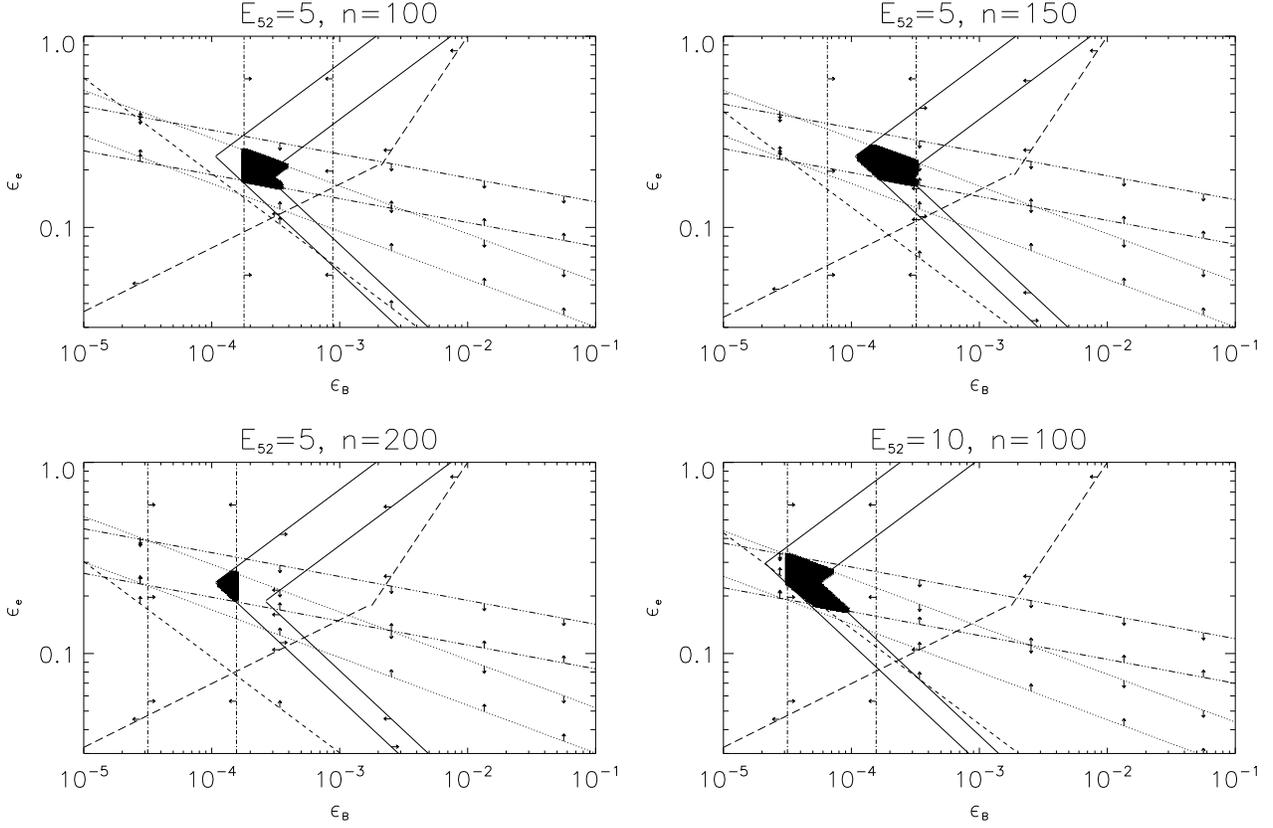}
\caption{The shadowed regions are the parts of the $\epsilon_{B}$-$\epsilon_{e}$ plane where the conditions expressed by Eq. (\ref{eq:cond_num}) (dotted lines), Eq. (\ref{eq:cond_nuc}) (dashed line), Eq. (\ref{eq:normx}) (solid lines),  Eq. (\ref{eq:cond_ficm}) (dash-dotted lines), Eqs. (\ref{eq:cond_numic1}) and (\ref{eq:cond_numic3}) (dash-dot-dot-dotted lines) are simultaneously satisfied, for $E_{52}=5$ and $n=100$ (upper-left panel), $E_{52}=5$ and $n=150$ (upper-right panel), $E_{52}=5$ and $n=200$ (lower-left panel), and $E_{52}=10$ and $n=100$ (lower-right panel). The long-dashed line marks the portion of the $\epsilon_{B}$-$\epsilon_{e}$ plane where $x=\left(\frac{\epsilon_{e}}{\epsilon_{B}}\right)^{1/2} > 10$ if $t^{IC}_{0} > 10^{6}$~s and $x (10^{6}~$s$)=\left(\frac{\epsilon_{e}}{\epsilon_{B}}\right)^{1/2}\left(\frac{10^{6}~{\rm s}}{t^{IC}_{0}}\right)^{-\frac{(p-2)}{2(4-p)}} > 10$ if $t^{IC}_{0} < 10^{6}$, so as to assure that $x > 1$ up to $10^{6}$~s, which is necessary for the consistency of
our formulation \citep{SariEsin2001}.} 
\label{diagrammi_rev_zoom}
\end{figure*}

To model the observed early-time emission within the synchrotron fireball model we choose scenario b) of Table \ref{tab:relazioni_chiusura} with $\nu_{c}$(100~s)$ \le 10^{14}$~Hz  and $\nu_{m}($100~s)~$\sim 0.2$~keV. We thus set $p=2.5$.

To model the flattening observed in the X-ray afterglow at late times, we add the contribution of an IC component. Following the prescriptions given by \citet{SariEsin2001}, the IC spectrum is modeled as a power-law spectrum similar to the synchrotron one, plus logarithmic corrections when relevant in the considered regime. In the power-law approximation, the IC spectrum is normalized to a peak-flux value of $f^{IC}_{max}=2~\times10^{-7}~f_{max}~n~(R/10^{18})$, where $f_{max}$ is the peak flux of the synchrotron component and $R$ is the fireball radius in cm \citep{SariEsin2001}.

To constrain the values of $E_{52}$, $n$, $\epsilon_{e}$, $\epsilon_{B}$ that reliably circumscribe the portion of parameter space compatible with the scenario we are testing, we set the following conditions:
\begin{equation}
0.1~{\rm keV} \lesssim \nu_{m}(100~\rm{s}) \lesssim 0.3~{\rm keV},
\label{eq:cond_num}
\end{equation}
\begin{equation}
\nu_{c}(100~{\rm s})\lesssim10^{14}~{\rm Hz},
\label{eq:cond_nuc}
\end{equation}
\begin{equation}
50 \times 10^{-3}~{\rm mJy} \lesssim f^{syn}_{0.2~{\rm keV}}(100~{\rm s}) \lesssim 70 \times 10^{-3}~{\rm mJy},
\label{eq:normx}
\end{equation}
where $f^{syn}_{0.2~{\rm keV}}(100~{\rm s})=f^{syn}_{max}\left(\frac{0.2~{\rm keV}}{\nu_{c}(100~{\rm s})}\right)^{-0.5}$ if $\nu_{m}(100~{\rm s}) > 0.2$~keV or $f^{syn}_{0.2~{\rm keV}}(100~{\rm s})=f^{syn}_{max}\left(\frac{\nu_{m}(100~{\rm s})}{\nu_{c}(100~{\rm s})}\right)^{-0.5}\left(\frac{0.2~{\rm keV}}{\nu_{m}(100~{\rm s})}\right)^{-p/2}$, if $\nu_{m}(100~{\rm s}) < 0.2$~keV (depending on Eq. (\ref{eq:cond_num}) and supposing that Eq. (\ref{eq:cond_nuc}) is valid).

Equations (\ref{eq:cond_num}) and (\ref{eq:cond_nuc}) take the requirement to overestimate as little as possible the early-time optical flux into account (see Sect. \ref{sec3.1}) and reproduce the observed X-ray spectral and temporal indices (for $p=2.5$, see Table1); Eq. (\ref{eq:normx}) is a normalization condition on the X-ray light curve based on the observed $0.2-10$~keV flux at $100$~s normalized to a frequency of $10^{18}$~Hz (at the center of the X-ray band). Moreover, to relate the observed flattening to the appearance of an IC component, the following conditions have to be satisfied:
\begin{equation}
4.5\times10^{-6}~{\rm mJy}\lesssim~f^{IC}_{max}(10^{5}~{\rm s})~\lesssim 10^{-5}~{\rm mJy},
\label{eq:cond_ficm}
\end{equation}
\begin{equation}
\nu^{IC}_{m}(10^{5}~{\rm s})~\gtrsim 5~{\rm keV},
\label{eq:cond_numic1}
\end{equation}
\begin{equation}
\nu^{IC}_{m}(10^{6}~{\rm s})~\lesssim0.2~{\rm keV}.
\label{eq:cond_numic3}
\end{equation}
\begin{figure*}[!]
\centering
\includegraphics{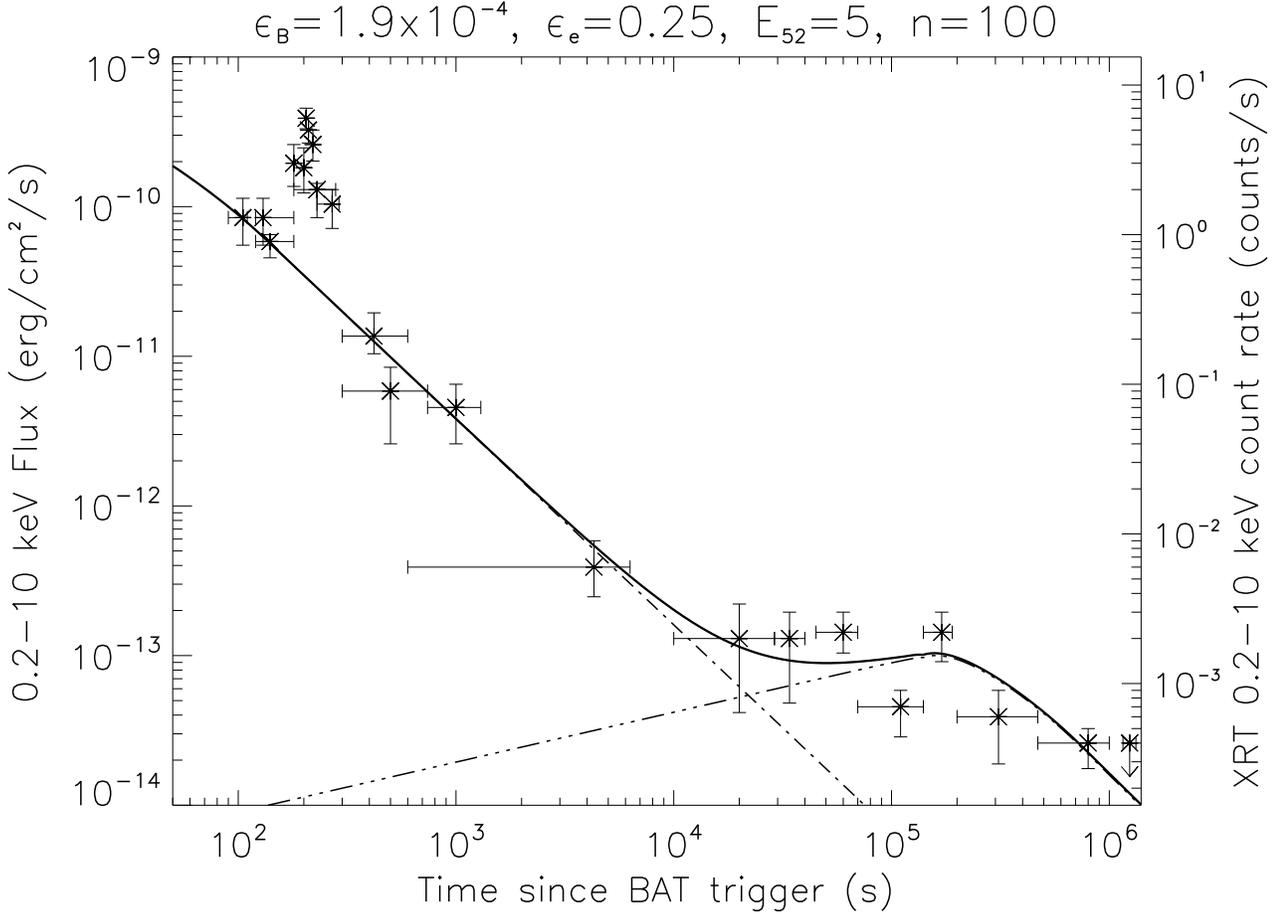}
\caption{Synchrotron-plus-IC model predictions for the $0.2-10$~keV light curve with $\epsilon_{B}=1.9\times10^{-4}$, $\epsilon_{e}=0.25$, $E_{52}=5$, $n=100$, $p=2.5$. The dash-dotted line represents the synchrotron component, while the dash-dot-dot-dotted line the IC one. The solid line is the resulting total flux. The model predictions are compared with the $0.2-10$~keV data points (crosses) by \citet{Romano2005}.}
\label{slow}
\end{figure*}

Considering that in both the fast and slow cooling regimes the IC light curve at a given frequency is expected to decline only after $\nu^{IC}_{m}$ becomes lower than that frequency, Eqs. (\ref{eq:cond_numic1}) and (\ref{eq:cond_numic3}) are set to reproduce the observed shape of the late-time flattening. In the $0.2-10$~keV band, the IC component should have a constant or rising profile up to $10^{5}$~s, while after that it should start decreasing, and around $10^{6}$~s it should no longer be visible above the background. We numerically tested for which values of $\epsilon_{B}$ and $\epsilon_{e}$ the above conditions are simultaneously satisfied, once the values of $E_{52}$ and $n$ are fixed. Figure \ref{diagrammi_rev_zoom} shows some examples on how the conditions expressed in Eqs. (\ref{eq:cond_num}) to (\ref{eq:cond_numic3}) determine the allowed range of parameter values. 

Among the possible combinations of parameters satisfying the conditions we set to determine the most promising portion of the parameter space, one should then specifically check for agreement with the actual data. In fact, in Fig. \ref{slow} we show how, for $\epsilon_{B}=1.9\times10^{-4}$, $\epsilon_{e}=0.25$, $E_{52}=5$, $n=100$, the appearance of an IC component can be a viable model to explain the late-time flattening observed in the X-ray light curve of \object{XRF~050406}.
\begin{figure}
\centering
\resizebox{\hsize}{!}{\includegraphics{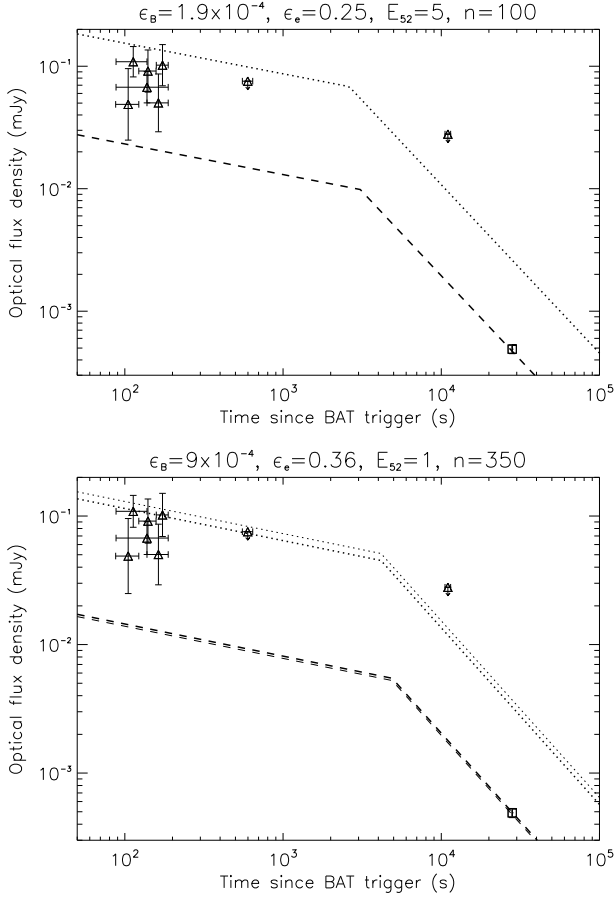}}
\caption{Synchrotron-plus-IC model predictions compared with the observed data in the case $\epsilon_{B}=1.9\times10^{-4}$, $\epsilon_{e}=0.25$, $E_{52}=5$, $n=100$, $p=2.5$ (upper panel) and  $\epsilon_{B}=9\times10^{-4}$, $\epsilon_{e}=0.36$, $E_{52}=1$, $n=350$, $p=2.5$ (lower panel). In the upper panel, the $V$-band (dotted line) and $R$-band (dashed line) light curves include a SMC-like local ($z=2.44$) extinction of $A(V_{int})\sim0.32$~mag; in the lower panel, the $V$-band (dotted lines) and $R$-band (dashed lines) light curves include an SMC-like extinction of $A(V_{int})\sim0.13$~mag (thick lines) or a Galactic-like $A(V_{int})\sim0.15$~mag (thin lines). In both panels the $R$-band light curve and data point have been shifted down by a factor of 10 for clarity.}
\label{ottico}
\end{figure}

For the same set of parameters, we also computed the predicted $V$- and $R$-band light curves. As expected (Sect. \ref{sec3.1}), if no local extinction is added, the predicted $V$- and $R$-band fluxes are overestimated of a factor of $\sim 5$ and $\sim 2.4$, respectively. In a ``Small Magellanic Clouds'' (SMC)-like environment \citep{Pei1992}, an intrinsic extinction of $A(V_{int})\sim0.32$~mag in the GRB site (at $z=2.44$) allows consistency to be recovered with the optical data (Fig. \ref{ottico}, upper panel); this implies an absorption column density of $N_{\rm{H}} \sim 0.32 \times 1.6 \times10^{22}$cm$^{-2}$. In such an environment, due to the lower metallicity, the upper-limit of $9\times10^{20}$~cm$^{-2}$ set by the XRT analysis should be increased by a factor of $\sim 7$ \citep{Stratta2004}, and thus the required $N_{\rm{H}}$ should be compared with this increased limit.
\begin{figure}
\centering
\resizebox{\hsize}{!}{\includegraphics{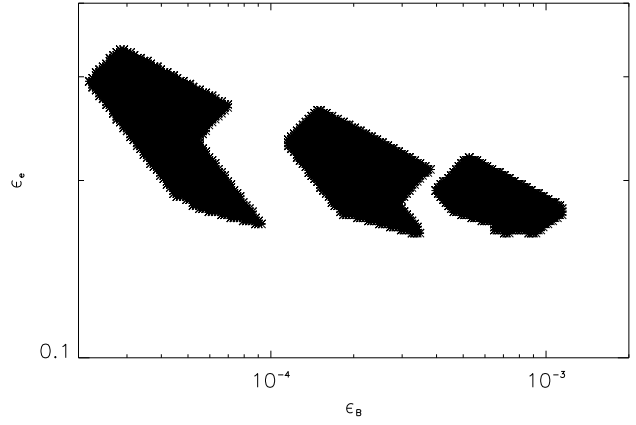}}
\caption{Numerical test results showing for which values of $\epsilon_{B}$ and $\epsilon_{e}$ the conditions expressed by Eqs. (\ref{eq:cond_num}) to (\ref{eq:cond_numic3}) are simultaneously satisfied (and $x > 10$ up to $10^{6}$~s). We repeated this search for $E_{52}=0.5,~1,~2,~3,~5,~10$ combined with $n=10,~50,~100,~150,~200,~250,~300,~350,~400$. Solutions are found for $E_{52}=3$ and $n=100,~150,~200$ (shadowed region on the right); $E_{52}=5$ and $n=100,~150,~200$ (shadowed region in the center); $E_{52}=10$ and $n=100,~150,~200$ (shadowed region on the left). The three regions represent, for each value of $E_{52}$, the superposition of regions similar to those represented in Fig. \ref{diagrammi_rev_zoom}, each corresponding to one value of $n$ among the ones allowed for the considered value of $E_{52}$. For example, the central region is the superposition of the three regions shown in the upper-left, upper-right, and lower-left panels of Fig. \ref{diagrammi_rev_zoom}.} 
\label{diagrammi_rev_def}
\end{figure}
The numerical test that guided us in finding this solution was repeated for $E_{52}=0.5,~1,~2,~3,~5,~10$, combined with $n=10,~50,~100,~150,~200,~250,~300,~350,~400$. Solutions are found for $E_{52}=3,~5,~10$ (Fig. \ref{diagrammi_rev_def}, right-to-left) and for $n=100,~150,~200$.
\begin{figure}
\centering
\resizebox{\hsize}{!}{\includegraphics{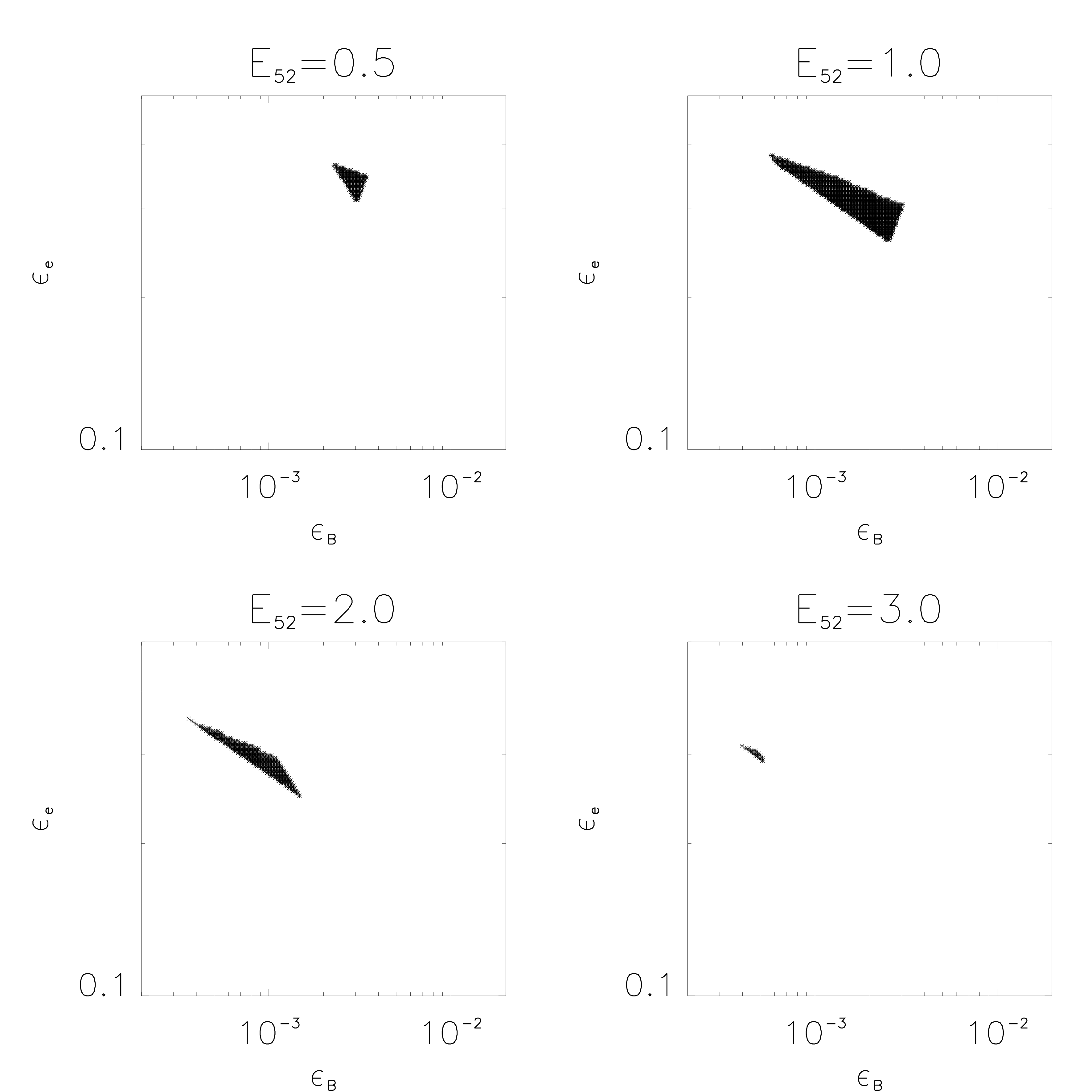}}
\caption{The shadowed regions are the parts of the $\epsilon_{B}$-$\epsilon_{e}$  plane where the required conditions (see text) are simultaneously satisfied (and $x > 10$ up to $10^{6}$~s). Solutions are found for $E_{52}=0.5$ and $n=350,~400$ (upper-left panel), $E_{52}=1$ and $n=200,~250,~300,~350,~400$ (upper-right panel), $E_{52}=2$ and $n=150,~200,~250,~300,~350$ (lower-left panel), and $E_{52}=3$ and $n=150,~200$ (lower-right panel).} \label{diagrammi_revF_def}
\end{figure}

If before the X-ray flare (say $t\lesssim~300$~s) there is some contribution from  the rising part of the flare itself (as one may expect in a late internal shock scenario), we can relax the normalization condition to some extent. Setting $\nu_{m}(300~s)\sim 0.2$~keV (so as to reproduce the observed X-ray spectral and temporal indices after the flare for $p=2.5$) and normalizing the synchrotron spectrum to fit the observed X-ray flux level at $\sim 300$~s, the  0.2~keV light curve at $t \lesssim 300$~s will rise slowly (as $t^{-1/4}$), and the 0.2~keV flux at $100$~s will be lowered by a factor of $\frac{(300~{\rm s}/100~{\rm s})^{-1/4}}{(300~{\rm s}/100~{\rm s})^{-3(p-1)/4-1/4}}\sim 3$ with respect to the previous solution (Fig. \ref{slow}). Extrapolating to the optical band, the expected $V$-band flux will overestimate the observed one by a factor of $\sim 5/3 \sim 2$, instead of $\sim 5$.

Under this hypothesis, we can shift the time in Eqs. (\ref{eq:cond_num}) and (\ref{eq:cond_nuc}) from $100$~s to $300$~s and modify the normalization condition on the synchrotron component (Eq. (\ref{eq:normx})) to fit the observed X-ray flux after the flare:
\begin{equation}
8\times10^{-3}~{\rm mJy}\lesssim~f^{syn}_{0.2~{\rm keV}}(300~{\rm s})\lesssim~20\times10^{-3}~{\rm mJy .}\end{equation}

\begin{figure*}[!]
\centering
\includegraphics{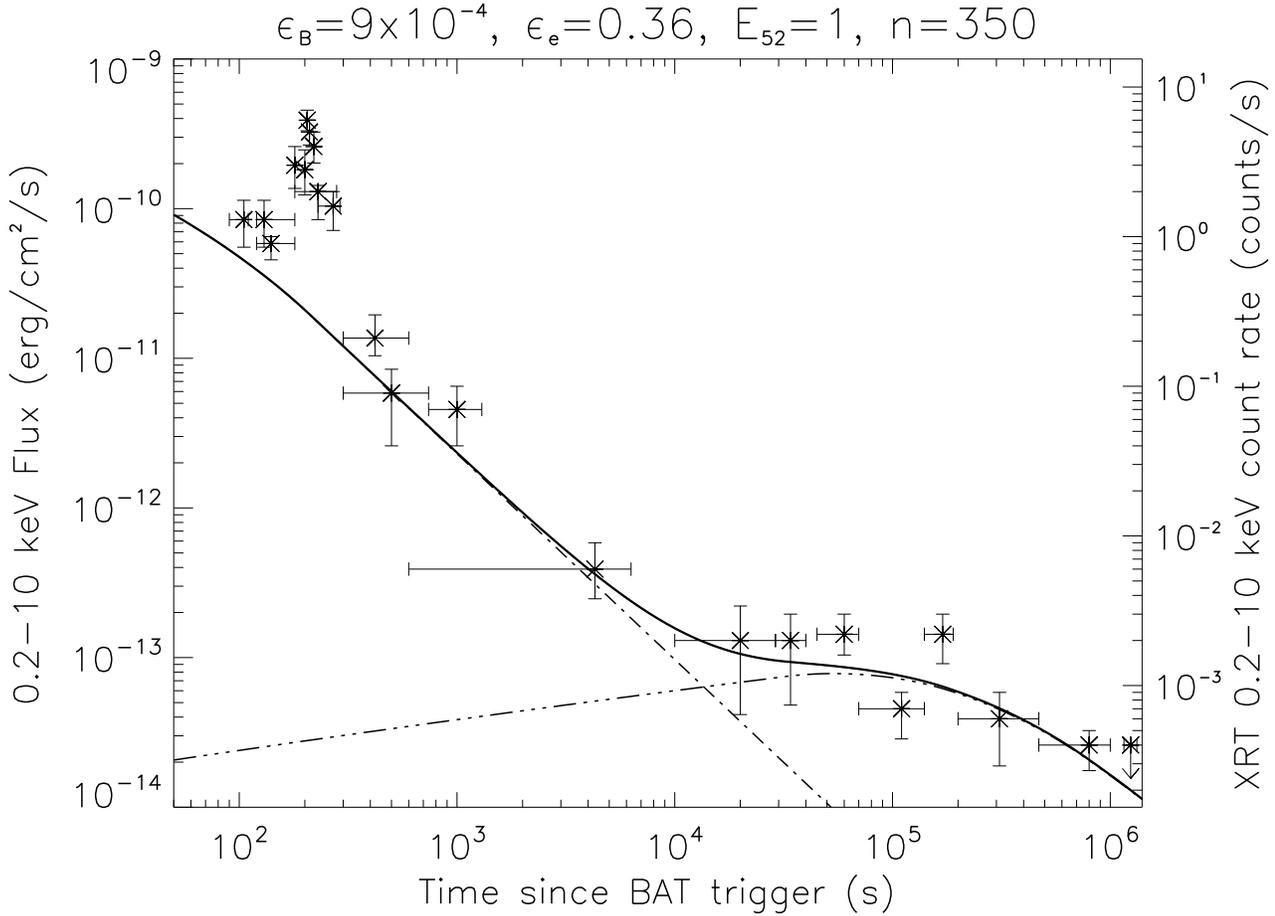}
\caption{Synchrotron plus IC model predictions for the $0.2-10$~keV light curve with $\epsilon_{B}=9\times10^{-4}$, $\epsilon_{e}=0.36$, $E_{52}=1$, $n=350$, $p=2.5$, compared with the observed data (crosses). The dash-dotted line is the contribution from the synchrotron component while the dash-dot-dot-dotted line the one from IC emission. The solid line is the resulting total flux.}
\label{fast}
\end{figure*}

Similar to what done before, we numerically searched for the values of $\epsilon_{B}$ and $\epsilon_{e}$ that allow us to satisfy simultaneously those conditions, for a given choice of $E_{52}$ and $n$. Solutions are found for $E_{52}=0.5, 1, 2, 3$ (see Fig. \ref{diagrammi_revF_def}) and values of $n$ between $100$ and $400$. In Fig. \ref{fast} we find a combination of parameter values, $\epsilon_{B}=9.0\times10^{-4}$, $\epsilon_{e}=0.36$, $E_{52}=1$, $n=350$ for which the predicted light curve agrees with the data. In this case the logarithmic corrections to the IC spectrum are significant between $0.2-10$ keV, and they have been added to the power-law approximation. The IC spectrum has been normalized to a peak-flux value of $f^{IC}_{max}=14/45~\times~f_{max}~\sigma_{\rm{T}}~n~R$ \citep{SariEsin2001}, where $\sigma_{\rm{T}}$ is the Thompson cross section.

In the optical band (Fig. \ref{ottico}, lower panel), the addition of a Galactic-like extinction term \citep[][ with $R_{V}=3.1$]{Cardelli1989}, with $A(V_{int})\sim 0.15$~mag in the GRB site ($z=2.44$), allows us to explain both the $R$- and $V$-band observations. A SMC-like environment with $A(V_{int})\sim 0.13$~mag could also be an alternative solution. The implied values of the local $N_{\rm{H}}$ are $\sim 0.15 \times 1.79 \times10^{21}$~cm$^{-2}$ and $\sim 0.13 \times 1.6 \times10^{22}$~cm$^{-2}$, both fully compatible with the upper-limit from the X-ray analysis. (For an SMC-like environment, as already noted, this upper-limit should be increased by a factor of $\sim 7$.)

We can finally focus on the spectral predictions of the model we are proposing to explain the XRF~050406 late-time flattening. In a synchrotron-plus-IC scenario, since the light curve flattening is associated with $\nu^{IC}_{m}$ crossing the X-ray band, the spectral index should vary between the values of $-1/3$ (for $t~\lesssim 10^{5}$~s the X-ray band is below $\nu^{IC}_{c}$) and $\sim (p-1)/2$ for the first solution proposed here (Fig. \ref{slow}) and between $0.5$ (for $2\times10^{4}$~s$~\lesssim t~\lesssim 2\times10^{5}$~s the X-ray band is between $\nu^{IC}_{c}$ and $\nu^{IC}_{m}$) and $p/2$ (plus logarithmic term corrections) for the second solution (Fig. \ref{fast}). The predicted mean spectral indices between $2.0\times10^4$~s and $10^{6}$~s are $<\beta>~\sim0.4$ and $<\beta>~\sim0.8$ for the first and second solutions, respectively. Comparing those values with the mean spectral index of $\beta=1.1\pm0.3$, measured before the flattening \citep{Romano2005}, it is evident that a hardening should be observed at late times.

For \object{XRF~050406}, about $60$ source counts were collected between $\sim2.0\times10^{4}$~s and $\sim10^{6}$~s, so that a detailed spectral analysis cannot be performed. A ratio of $1.8\pm0.5$ between hard ($H: 0.2-1.0$~keV) and soft ($S: 1.0-10$~keV) photons in the whole interval time is the only information that one can get from this faint source (private communication by P. Romano, 2006). Comparing this with $0.5 \lesssim H/S \lesssim 1.0$ obtained by \citet{Romano2005} after the flare \citep[see the lowest panel in Fig. 3 of ][]{Romano2005}, the spectrum during the late-time flattening seems to be harder, but no firm conclusion can be reached due to the large errors. Since hardening is a natural expectation in a synchrotron-plus-IC scenario, we then investigated on the compatibility of our predictions with the observed $H/S$. Using the XRT PC response\footnote{http://heasarc.nasa.gov/Tools/w3pimms.html} (Grade 0-12, on-axis counts, infinite extraction region), assuming $N_{H}=2.8\times10^{20}$~cm$^{-2}$, and normalizing the spectrum so as to have about $60$ counts in total, we computed a rough estimate of the expected H/S ratio for a spectrum with $<\beta>~\sim 0.4$ and $<\beta>~\sim 0.8$, obtaining $H/S=2.4\pm0.7$ and $H/S=1.5\pm0.4$ for the two spectra\footnote{To check for consistency, we also computed the expected hardness ratio for $\beta=1.1$ and $100$ source counts. We obtained $H/S=0.98\pm0.19$, consistent with the values of $H/S$ found by \citet{Romano2005} after the flare.}; those values of $H/S$ are both compatible with the observed value of $1.8\pm0.5$ within the errors. Although a conclusive statement is not allowed, encouraging indications in favor of the present model as a mechanism for the late-time flattening arise globally.

We thus conclude that future observations of brighter sources, possibly allowing performance of time-resolved spectroscopy during the late-time flattening, will be a key to confirming or rejecting the scenario we are suggesting and its potential extension as a general explanation for this kind of late-time behavior in GRB X-ray light curves.

\section{Conclusions}
\label{sec4} We discussed the \object{XRF~050406} afterglow in the context of the standard fireball model. Within a synchrotron-plus-IC scenario, we tested whether the late-time flattening observed in the X-ray light curve can be explained by the appearance of an IC component. We found that setting $\epsilon_{B}=1.9\times10^{-4}$, $\epsilon_{e}=0.25$, $E_{52}=5$, and $n=100$ can explain the X-ray observations. Considering the optical data as well, we noted that the early optical-to-X-ray spectral index appears rather flat and requires additional extinction to recover consistency with the data. We also proposed a second solution with $\epsilon_{B}=9\times10^{-4}$, $\epsilon_{e}=0.36$, $E_{52}=1$, $n=350$, where the optical-to-X-ray normalization problem is solved requiring that at $t < 210$~s some contribution to the X-ray emission comes from the rising part of the flare. Using a Galactic- or SMC-like extinction curve, the $N_{\rm{H}}$ required by the optical data is consistent with the upper-limit found in the X-ray analysis.
 \begin{acknowledgements}
L. Piro acknowledges the support of the EU through the EU FPC5 RTN ``Gamma-ray burst, an enigma and a tool''; A. Corsi acknowledges the support of an INFN grant. We thank Pat Romano for performing the spectral analysis of the late-time data and providing the information necessary to test an important aspect of the present work; A. Corsi would also like to thank Pat Romano for interesting general discussions, Giovanni Montani for important comments and suggestions, and Patricia Schady for useful discussions of the optical data.
\end{acknowledgements}
\bibliographystyle{aa}
\bibliography{Corsi}
\end{document}